\documentclass[runningheads]{llncs}
\usepackage[T1]{fontenc}
\usepackage{graphicx}
\begin{document}

\newcommand{\repeatthanks}{\textsuperscript{\thefootnote}}

\title{Brain Imaging Generation with Latent Diffusion Models}

\author{Walter H. L. Pinaya \inst{1}\thanks{Equal contribution} \and
Petru-Daniel Tudosiu \inst{1}\repeatthanks \and
Jessica Dafflon \inst{2,3} \and
Pedro F Da Costa \inst{4,5} \and
Virginia Fernandez \inst{1} \and
Parashkev Nachev \inst{6} \and
Sebastien Ourselin \inst{1} \and
M. Jorge Cardoso \inst{1}}

\authorrunning{Pinaya et al.}
\institute{Department of Biomedical Engineering, School of Biomedical Engineering \& Imaging Sciences, King's College London, UK \and
Data Science and Sharing Team, Functional Magnetic Resonance Imaging Facility, National Institute of Mental Health, Bethesda, MD, 20892, USA \and
Machine Learning Team, Functional Magnetic Resonance Imaging Facility, National Institute of Mental Health, Bethesda, MD, 20892, USA \and
Institute of Psychiatry, Psychology \& Neuroscience, King's College London, UK \and
Centre for Brain and Cognitive Development, Birkbeck College, UK \and
Institute of Neurology, University College London, UK}

\maketitle              
\begin{abstract}
Deep neural networks have brought remarkable breakthroughs in medical image analysis. However, due to their data-hungry nature, the modest dataset sizes in medical imaging projects might be hindering their full potential. Generating synthetic data provides a promising alternative, allowing to complement training datasets and conducting medical image research at a larger scale. Diffusion models recently have caught the attention of the computer vision community by producing photorealistic synthetic images. In this study, we explore using Latent Diffusion Models to generate synthetic images from high-resolution 3D brain images. We used T1w MRI images from the UK Biobank dataset (N=31,740) to train our models to learn about the probabilistic distribution of brain images, conditioned on covariables, such as age, sex, and brain structure volumes. We found that our models created realistic data, and we could use the conditioning variables to control the data generation effectively. Besides that, we created a synthetic dataset with 100,000 brain images and made it openly available to the scientific community.

\end{abstract}
\section{Introduction}

Deep neural networks fuelled several ground-breaking advancements in areas such as natural language processing and computer vision, where part of these improvements was attributed to the large amount of rich data used to train these networks, with some public datasets reaching millions of images and text sentences \cite{deng2009imagenet,schuhmann2021laion}. During the same period, medical image analysis also made remarkable breakthroughs by applying deep neural networks to solve tasks such as segmentation, structure detection, and computer-aided diagnosis (detailed review available at \cite{lundervold2019overview,shen2017deep}). However, one current limitation of medical imaging projects is the lack of availability of large datasets. Medical data are costly and laborious to collect, and privacy concerns create challenges to data sharing by restricting publicly available medical datasets to up to a few thousand examples. This limitation creates a bottleneck on models’ generalizability and hampers the rate at which cutting-edge methods are deployed in the clinical routine.

Generating synthetic data with privacy guarantees provides a promising alternative, allowing meaningful research to be carried out at scale  \cite{jordon2020synthetic,jordon2022synthetic,wang2021review}. Together with traditional data augmentation techniques (e.g., geometric transformations), these synthetic data could complement real data to dramatically increase the training set of machine learning models. Generative models learn the probability density function underlying the data they are trained on and can create realistic representations of examples which are different from the ones present in the training data by sampling from the learned distribution. However, generating meaningful synthetic data is not easy, especially when considering complex organs like the brain. 

Nowadays, Generative Adversarial Networks (GANs) have been applied in various fields to create synthetic images, producing realistic and clear images and achieving impressive performance \cite{creswell2018generative,wang2020state}. In the medical field, for example, \cite{kwon2019generation} combine variational autoencoders with GANs to generate various modalities of whole brain volumes from a small training set and achieved a better performance compared to several baselines. However, since their study resized the images to a small volume before training, with a size of 64 × 64 × 64 voxels, their synthetic medical images did not replicate many essential finer details. In addition, due to the prevalence of 3D high-resolution data in the field, researchers tend to have their models restrained by the amount of GPU memory available. To mitigate this problem, \cite{sun2022hierarchical} proposed a 3D GAN with a hierarchical structure which is able to generate a low-resolution version of the image and anchor the generation of high-resolution sub-volumes on it. With this approach, the authors were able to generate impressive realistic 3D thorax CT and brain MRI with resolutions up to 256 × 256 × 256 voxels. Despite generating great interest, GANs still come with inherent challenges, such as being notoriously unstable during training and failing to converge or to capture the variability of the generated data due to mode collapse issues \cite{kodali2017convergence}.

Recently, diffusion models caught the attention of the machine learning community by showing promising results when synthesizing natural images. They have rivalled GANs in sample quality \cite{dhariwal2021diffusion} while building upon a solid theoretical foundation. Not only have they reported impressive photorealism unconditioned images, but they have also been used to create images conditioned in classes and text sentences (using techniques like classifier-free guidance  \cite{ho2021classifier}), with exceptional results on models like Latent Diffusion Models \cite{rombach2022high}, DALLE 2 \cite{ramesh2022hierarchical}, and Imagen \cite{saharia2022photorealistic}. 

In this study, we used diffusion models to create synthetic MRI images of the adult human brain. For that, we used 31,740 training images from the UK Biobank \cite{sudlow2015uk}to train our models. In order to efficiently scale the application of diffusion models to these high-resolution 3D data, we combined our diffusion models with compression models following the architecture of Latent Diffusion Models (LDM) \cite{rombach2022high}. Furthermore, we conditioned the image generations on age, gender, ventricular volume, and brain volume relative to the intracranial volume in order to generate realistic examples of brain scans with specific covariate values. We compared our synthetic images to state-of-the-art methods based on GANs, and we made our synthetic dataset comprising 100,000 brain images publicly available to the scientific community.

\begin{figure}[t!]
\centering
\includegraphics[width=\textwidth]{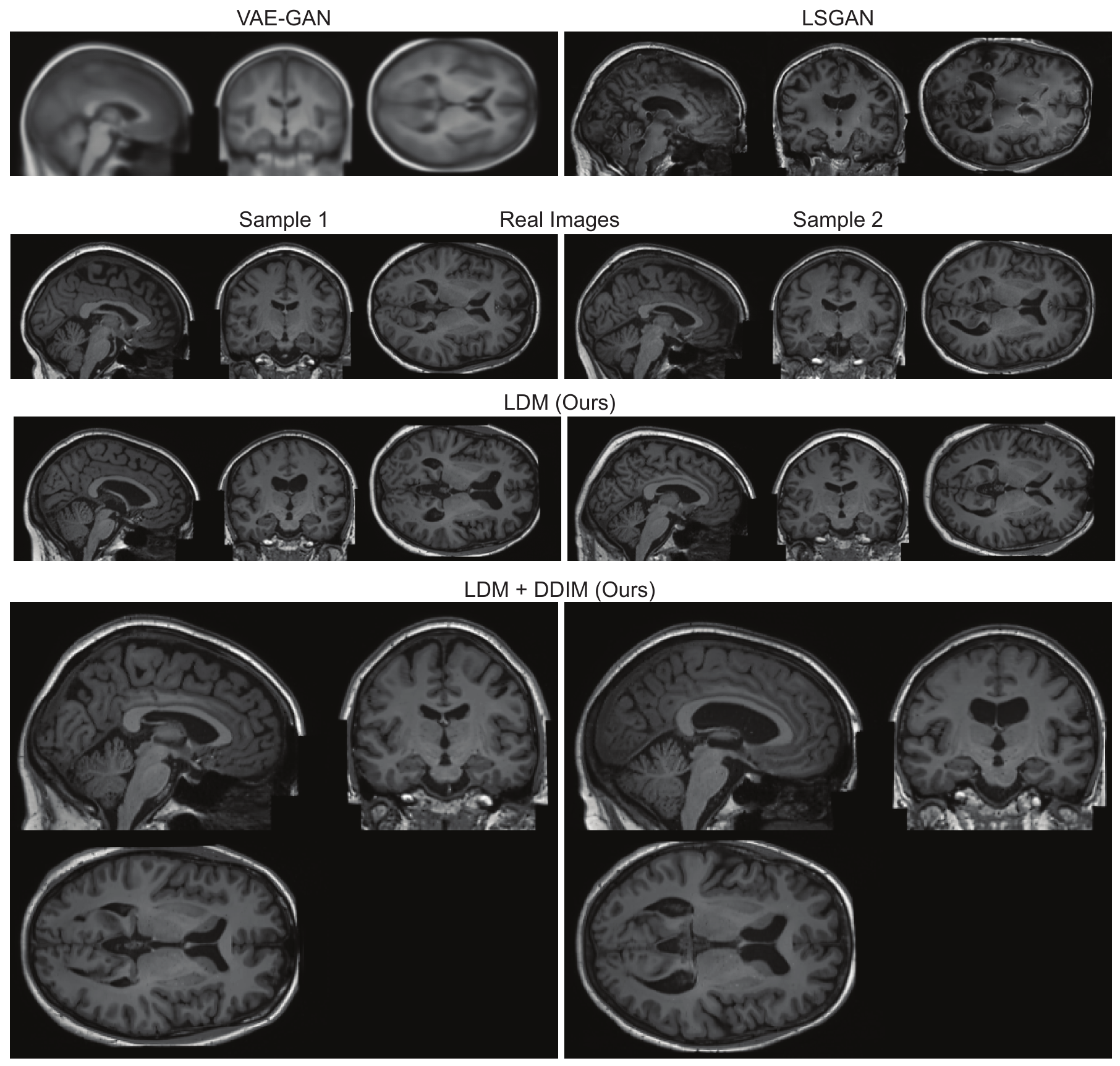}
\caption{Real and synthetic samples of head MRI generated using VAE-GAN, LSGAN, LDM and LDM+DDIM.} \label{fig1}
\end{figure}

\section{Methods}

\subsection{Datasets and Image Preprocessing}
In this study, we used images from the UK Biobank (UKB) \cite{sudlow2015uk} to train our generative models. The UKB is a study that aims to follow the health and well-being of volunteer participants across the United Kingdom. Here, we used an early release of the project's data comprising 31,740 participants with T1w images. The dataset consists of healthy individuals aged between 44 and 82 years with average age of 63.6 $\pm$ 7.5 years (average $\pm$ SD) and 14,942 male subjects (47\%). In our experiments, we also conditioned for the volume of ventricular cerebrospinal fluid (min-max: 6995.68 - 171375.0 $mm^3$; UKB Data-Field 25004) and brain volume normalised for head size (min-max: 1144240 - 1793910 $mm^3$; UKB Data-Field 25009). All variables used for model conditioning were normalised using min-max normalisation before feeding them to our models.

For the image pre-processing, we used UniRes\footnote[1]{https://github.com/brudfors/UniRes} \cite{brudfors2018mri,brudfors2019tool} to perform a rigid body registration to a common MNI space. The final images had 1 $mm^3$ as voxel size, and we cropped the image to obtain a volume of the head measuring 160 × 224 × 160 voxels.

\subsection{Generative models} 

In our experiments, we used LDMs, which combine the use of autoencoders to compress the input data into a lower-dimensional latent representation with the generative modelling properties of diffusion models. The compression model was an essential step to allow us to scale to high-resolution medical images. We trained the autoencoder with a combination of L1 loss, perceptual loss \cite{zhang2018unreasonable}, a patch-based adversarial objective \cite{esser2021taming}, and a KL regularization of the latent space. The encoder maps the brain image to a latent representation with a size of 20 × 28 × 20. After training the compression model, the latent representations of the training set are used as input to the diffusion model. Diffusion models \cite{ho2020denoising,sohl2015deep} are generative models that convert Gaussian noise into samples from a learned data distribution via an iterative denoising process. Given a latent representation of an example from our training set, the diffusion process gradually destroys the structure of the data via a fixed Markov chain over 1000 steps by adding Gaussian noise using a fixed linear variance schedule. The reverse process is also modelled as a Markov chain which learns to recover the original input from the noisy one. We conditioned our models according to age, gender, ventricular volume, and brain volume relative to the intracranial volume. To perform this conditioning, we used a hybrid approach combining the concatenation of the conditioning with the input data and the use of cross-attention mechanisms, as proposed in  \cite{rombach2022high}. Training and model details are available in the supplementary material.

\section{Experiments}

\subsection{Sampling Quality}

Fig.~\ref{fig1} shows images generated using LDMs compared to real images and the baselines (i.e., VAE-GAN \cite{larsen2016autoencoding} and LSGAN \cite{mao2017least}). Unlike the baselines, we observe that the LDMs were able to sample high-quality images with sharp details and realistic textures. Besides that, training the diffusion models at such a high resolution was much more stable and easier to achieve convergence when compared to the GAN-based baselines. The baselines required a meticulous design of the interaction between discriminator and generator, and they presented problems of mode collapse, showcasing the problems of GAN-based applied in such high-resolution 3D images. Therefore, we will refine and expand our comparisons with other baselines in future works.

We also obtained quantitative metrics about the performance of our models. We used the Fréchet Inception Distance (FID) \cite{heusel2017gans}to measure how realistic the synthetic images are. A small FID indicates that the distribution of the generated images is similar to the distribution of the real images. The FID was calculated using an approach similar to \cite{sun2022hierarchical}, where features were extracted using a pre-trained Med3D \cite{chen2019med3d}. We also measured the generation diversity with the Multi-Scale Structural Similarity Metric (MS-SSIM) and 4-G-R-SSIM \cite{rouse2008analyzing,chen2006gradient,li2010content}, where a value close to 0 suggests high diversity. Here, we presented the MS-SSIM for comparison with previous studies, but we also added the 4-G-R-SSIM as it has been shown to have better image quality assessment. We compute the average values from 1000 sample pairs. Table~\ref{tabquant} shows the quantitative results for different models used for the image synthesis.

\begin{table}[h!]
\caption{Quantitative evaluation of the synthetic images on the UK Biobank. We used the Fréchet Inception Distance (FID) to verify how realistic are the images and the multi-scale structural similarity metric (MS-SSIM) and 4-G-R-SSIM to evaluate generation diversity. We used 50 timesteps when sampling our models with DDIM sampler.}\label{tabquant}
\begin{center}
\begin{tabular}{lccc}
  \hline
   & \bfseries FID $\downarrow$ & \bfseries MS-SSIM $\downarrow$ & \bfseries 4-G-R-SSIM $\downarrow$  \\
  \hline
  \hline
  LSGAN             & 0.0231    & 0.9997    & 0.9969 \\
  VAE-GAN           & 0.1576    & 0.9671    & 0.8719 \\
  \hline
  LDM               & \bfseries 0.0076    & \bfseries 0.6555    & \bfseries 0.3883 \\
  LDM + DDIM        & 0.0080     & 0.6704     & 0.3957 \\
  \hline
  Real images       & 0.0005     & 0.6536     & 0.3909 \\
  \hline
\end{tabular}
\end{center}
\end{table}

Recently, different methods have been proposed to speed up the reverse process (e.g., Denoising Diffusion Implicit Models - DDIM), reducing by 10×~50× the number of necessary reverse steps \cite{song2020denoising}. Using the DDIM sampler, we reduced the number of timesteps from 1000 steps to only 50. This improved our sampling time from an average of 142.3$\pm$1.6s per sample to 7.6$\pm$0.2 s per sample @ NVIDIA TITAN RTX with minimum loss in performance (Table~\ref{tabquant}). Because of this boost in processing time and a minimal performance loss, we are using the LDM with the DDIM sampler for all the remaining analyses.

\begin{figure}[h!]
\centering
\includegraphics[width=\textwidth]{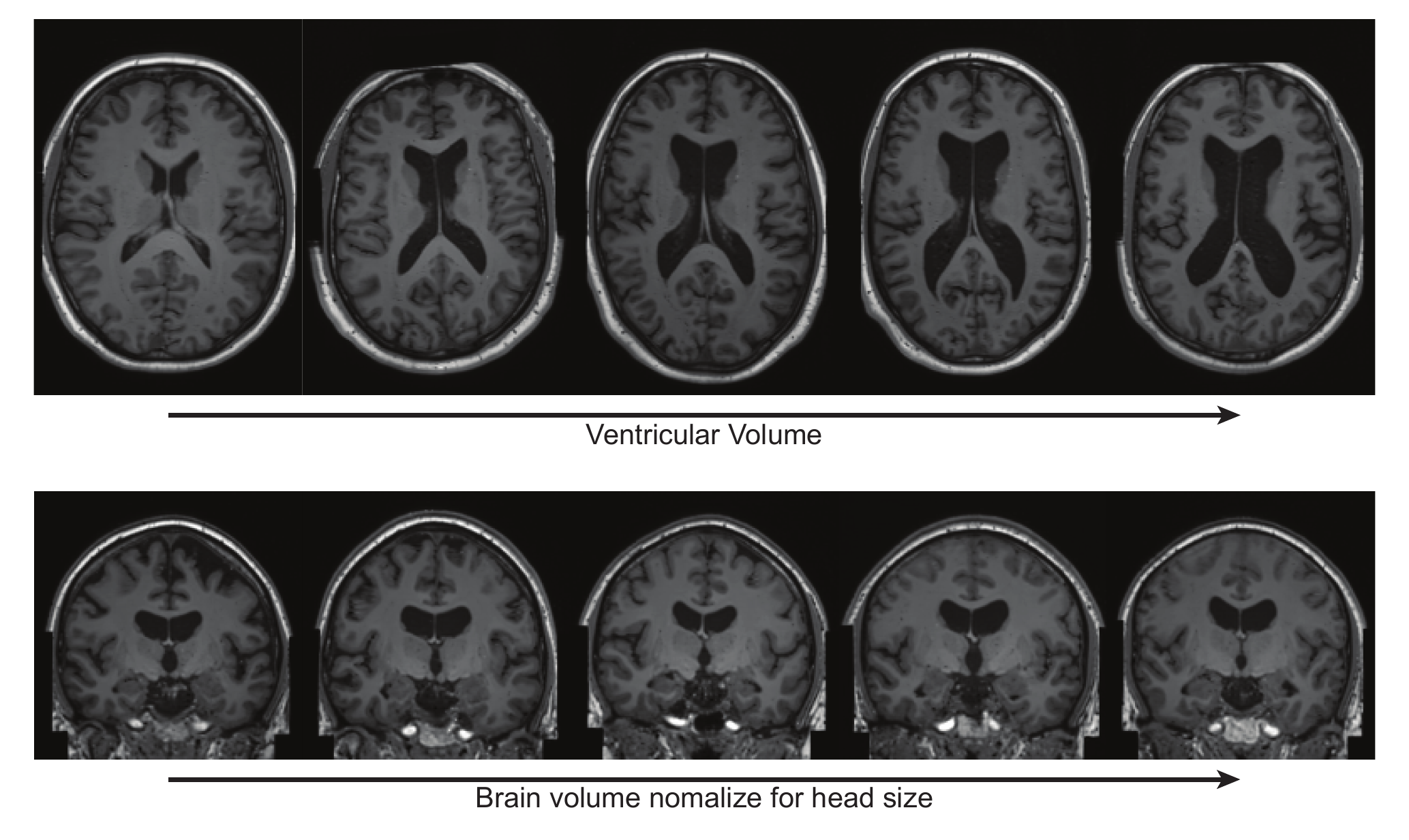}
\caption{Conditioned sampling varying the ventricular volume and the brain volume normalized by the intracranial volume. In both rows, we kept the other variables constant.} \label{fig2}
\end{figure}

\subsection{Conditioning Evaluation}
Using the hybrid conditioning approach \cite{rombach2022high}, we were able to condition our models and generate brain images where we can specify the age, sex, ventricular volume, and brain volume. As we can observe in Fig.~\ref{fig2}, our model was able to learn representations conditioned on regional (i.e., ventricular volume) and global (i.e., brain volume) volumes.

\begin{figure}[h!]
\centering
\includegraphics[width=\textwidth]{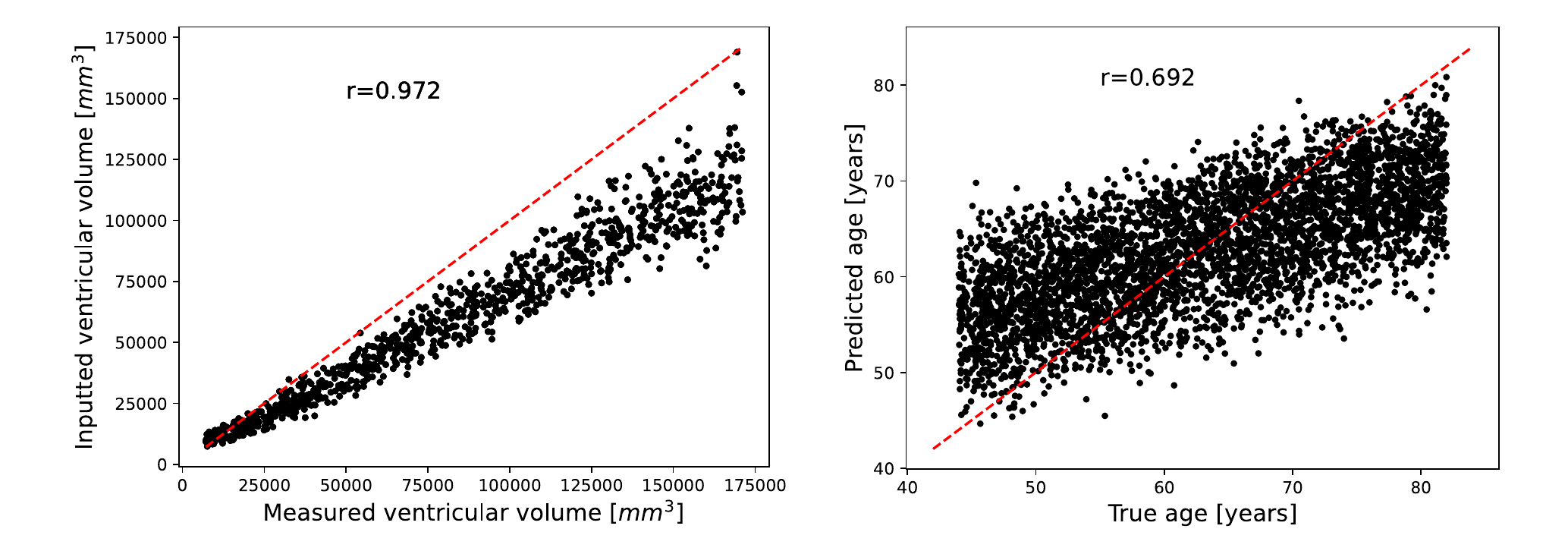}
\caption{Conditioning analysis. Left) Correlation between inputted ventricular volume vs ventricular measured with SynthSeg. Right) Correlation between inputted age and predicted brain age.} \label{fig3}
\end{figure}

In order to quantitatively evaluate the conditioning, we used SynthSeg \footnote[2]{https://github.com/BBillot/SynthSeg} \cite{billot2021synthseg} to measure the volumes of the ventricles of 1000 synthetic brains. In this analysis, we measured the combination of the left and right lateral ventricles and the left and right inferior lateral ventricles. We then computed the Pearson correlation between the obtained volumes and the inputted conditioning values. Using this approach, we observed a high correlation coefficient of 0.972, which demonstrates the effectiveness of conditioning on our model (Fig.~\ref{fig3}).

Additionally, we verified how well the effectiveness of conditioning brain generation by age. To this end, we used discriminative models to perform the task of brain age prediction, where we predict chronological age based on the brain image. In our study, we used the 3D convolution neural network proposed in \cite{cole2017predicting}, trained on the same training set used in the LDM training. After training the model, we verify how well the predicted age approximated the inputted age of the synthetic dataset. As shown in Fig.~\ref{fig3}, our model presented a high correlation between the inputted conditioning and the predicted age (r=0.692).

\begin{figure}[h!]
\centering
\includegraphics[width=\textwidth]{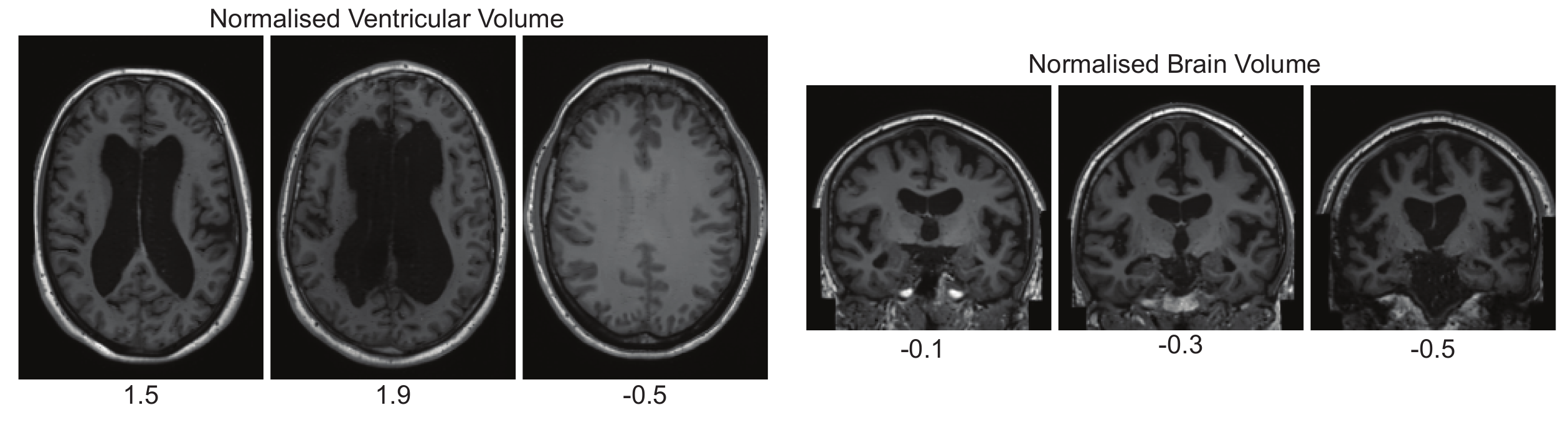}
\caption{Extrapolating values of conditioning variables. During the training of the models, the inputted values of the conditioning variables were scaled between 0 and 1. In this experiment, we tried values outside of this range, and we observed that our model could extrapolate the representation of brain and ventricular volumes, showing that it learned the concept of these variables.} \label{fig4}
\end{figure}

Finally, we verified how our model extrapolates the conditioning variables for values never shown during training. Fig.~\ref{fig4} presents samples where we used a normalised ventricular value higher than 1; in this case, we can see abnormally huge ventricles when using values of 1.5 and 1.9. If we use a negative value (e.g., -0.5), an image without ventricles is generated. Similarly, if we use negative values for the brain normalised for head size, the brain exhibits signs of neurodegeneration, showing smaller volumes of white and grey matter. These findings suggest that our models learned the concepts behind these conditioning variables during training.

\subsection{Synthetic Dataset}
We made a synthetic dataset of 100,000 human brain images generated by our model openly available to the community. This dataset is available at Academics Torrents\footnote[3]{https://academictorrents.com/details/63aeb864bbe2115ded0aa0d7d36334c026f0660b},  FigShare\footnote[4]{https://figshare.com/}, and HDRUK Gateway\footnote[5]{https://www.healthdatagateway.org/}, together with the conditioning information.

\section{Conclusions}
In our study, we were able to train diffusion models to effectively generate synthetic brain images that replicate properties from the training images. As it is the case with natural image generation, our diffusion models outperform alternative GANs-based methods in an unconditioned scenario. Additionally, we demonstrated how our methods could be conditioned on covariates such as age, sex, and brain structure volumes to produce the expected representation. In future works, we will develop models that use other scanning modalities as conditioning, such as images and radiological reports. By making the synthetic dataset openly available, this work also addresses one of the biggest limitations in medical machine learning – the challenge of obtaining large imaging datasets – while not posing threats to privacy infringements. In sum, our results show that LDMs are promising models to be explored in medical image generation.

\subsubsection{Acknowledgements}
WHLP and MJC are supported by Wellcome Innovations [WT213038/Z/18/Z]. PTD is supported by the EPSRC Research Council, part of the EPSRC DTP, grant Ref: [EP/R513064/1]. JD is supported by the Intramural Research Program of the NIMH (ZIC-MH002960 and ZIC-MH002968). PFDC is supported by the European Union’s HORIZON 2020 Research and Innovation Programme under the Marie Sklodowska-Curie Grant Agreement No 814302. PN is supported by Wellcome Innovations [WT213038/Z/18/Z] and the UCLH NIHR Biomedical Research Centre. This research has been conducted using the UK Biobank Resource (Project number: 58292).

\bibliographystyle{splncs04}
\bibliography{mybibliography}

\begin{thebibliography}{10}
\providecommand{\url}[1]{\texttt{#1}}
\providecommand{\urlprefix}{URL }
\providecommand{\doi}[1]{https://doi.org/#1}

\bibitem{billot2021synthseg}
Billot, B., Greve, D.N., Puonti, O., Thielscher, A., Van~Leemput, K., Fischl,
  B., Dalca, A.V., Iglesias, J.E.: Synthseg: Domain randomisation for
  segmentation of brain mri scans of any contrast and resolution. arXiv
  preprint arXiv:2107.09559  (2021)

\bibitem{brudfors2018mri}
Brudfors, M., Balbastre, Y., Nachev, P., Ashburner, J.: Mri super-resolution
  using multi-channel total variation. In: Annual Conference on Medical Image
  Understanding and Analysis. pp. 217--228. Springer (2018)

\bibitem{brudfors2019tool}
Brudfors, M., Balbastre, Y., Nachev, P., Ashburner, J.: A tool for
  super-resolving multimodal clinical mri. arXiv preprint arXiv:1909.01140
  (2019)

\bibitem{chen2006gradient}
Chen, G.H., Yang, C.L., Xie, S.L.: Gradient-based structural similarity for
  image quality assessment. In: 2006 international conference on image
  processing. pp. 2929--2932. IEEE (2006)

\bibitem{chen2019med3d}
Chen, S., Ma, K., Zheng, Y.: Med3d: Transfer learning for 3d medical image
  analysis. arXiv preprint arXiv:1904.00625  (2019)

\bibitem{cole2017predicting}
Cole, J.H., Poudel, R.P., Tsagkrasoulis, D., Caan, M.W., Steves, C., Spector,
  T.D., Montana, G.: Predicting brain age with deep learning from raw imaging
  data results in a reliable and heritable biomarker. NeuroImage  \textbf{163},
   115--124 (2017)

\bibitem{creswell2018generative}
Creswell, A., White, T., Dumoulin, V., Arulkumaran, K., Sengupta, B., Bharath,
  A.A.: Generative adversarial networks: An overview. IEEE signal processing
  magazine  \textbf{35}(1),  53--65 (2018)

\bibitem{deng2009imagenet}
Deng, J., Dong, W., Socher, R., Li, L.J., Li, K., Fei-Fei, L.: Imagenet: A
  large-scale hierarchical image database. In: 2009 IEEE conference on computer
  vision and pattern recognition. pp. 248--255. Ieee (2009)

\bibitem{dhariwal2021diffusion}
Dhariwal, P., Nichol, A.: Diffusion models beat gans on image synthesis.
  Advances in Neural Information Processing Systems  \textbf{34},  8780--8794
  (2021)

\bibitem{esser2021taming}
Esser, P., Rombach, R., Ommer, B.: Taming transformers for high-resolution
  image synthesis. In: Proceedings of the IEEE/CVF conference on computer
  vision and pattern recognition. pp. 12873--12883 (2021)

\bibitem{heusel2017gans}
Heusel, M., Ramsauer, H., Unterthiner, T., Nessler, B., Hochreiter, S.: Gans
  trained by a two time-scale update rule converge to a local nash equilibrium.
  Advances in neural information processing systems  \textbf{30} (2017)

\bibitem{ho2020denoising}
Ho, J., Jain, A., Abbeel, P.: Denoising diffusion probabilistic models.
  Advances in Neural Information Processing Systems  \textbf{33},  6840--6851
  (2020)

\bibitem{ho2021classifier}
Ho, J., Salimans, T.: Classifier-free diffusion guidance. In: NeurIPS 2021
  Workshop on Deep Generative Models and Downstream Applications (2021)

\bibitem{jordon2022synthetic}
Jordon, J., Szpruch, L., Houssiau, F., Bottarelli, M., Cherubin, G., Maple, C.,
  Cohen, S.N., Weller, A.: Synthetic data--what, why and how? arXiv preprint
  arXiv:2205.03257  (2022)

\bibitem{jordon2020synthetic}
Jordon, J., Wilson, A., van~der Schaar, M.: Synthetic data: Opening the data
  floodgates to enable faster, more directed development of machine learning
  methods. arXiv preprint arXiv:2012.04580  (2020)

\bibitem{kodali2017convergence}
Kodali, N., Abernethy, J., Hays, J., Kira, Z.: On convergence and stability of
  gans. arXiv preprint arXiv:1705.07215  (2017)

\bibitem{kwon2019generation}
Kwon, G., Han, C., Kim, D.s.: Generation of 3d brain mri using auto-encoding
  generative adversarial networks. In: International Conference on Medical
  Image Computing and Computer-Assisted Intervention. pp. 118--126. Springer
  (2019)

\bibitem{larsen2016autoencoding}
Larsen, A.B.L., S{\o}nderby, S.K., Larochelle, H., Winther, O.: Autoencoding
  beyond pixels using a learned similarity metric. In: International conference
  on machine learning. pp. 1558--1566. PMLR (2016)

\bibitem{li2010content}
Li, C., Bovik, A.C.: Content-partitioned structural similarity index for image
  quality assessment. Signal Processing: Image Communication  \textbf{25}(7),
  517--526 (2010)

\bibitem{lundervold2019overview}
Lundervold, A.S., Lundervold, A.: An overview of deep learning in medical
  imaging focusing on mri. Zeitschrift f{\"u}r Medizinische Physik
  \textbf{29}(2),  102--127 (2019)

\bibitem{mao2017least}
Mao, X., Li, Q., Xie, H., Lau, R.Y., Wang, Z., Paul~Smolley, S.: Least squares
  generative adversarial networks. In: Proceedings of the IEEE international
  conference on computer vision. pp. 2794--2802 (2017)

\bibitem{ramesh2022hierarchical}
Ramesh, A., Dhariwal, P., Nichol, A., Chu, C., Chen, M.: Hierarchical
  text-conditional image generation with clip latents. arXiv preprint
  arXiv:2204.06125  (2022)

\bibitem{rombach2022high}
Rombach, R., Blattmann, A., Lorenz, D., Esser, P., Ommer, B.: High-resolution
  image synthesis with latent diffusion models. In: Proceedings of the IEEE/CVF
  Conference on Computer Vision and Pattern Recognition. pp. 10684--10695
  (2022)

\bibitem{rouse2008analyzing}
Rouse, D.M., Hemami, S.S.: Analyzing the role of visual structure in the
  recognition of natural image content with multi-scale ssim. In: Human Vision
  and Electronic Imaging XIII. vol.~6806, pp. 410--423. SPIE (2008)

\bibitem{saharia2022photorealistic}
Saharia, C., Chan, W., Saxena, S., Li, L., Whang, J., Denton, E., Ghasemipour,
  S.K.S., Ayan, B.K., Mahdavi, S.S., Lopes, R.G., et~al.: Photorealistic
  text-to-image diffusion models with deep language understanding. arXiv
  preprint arXiv:2205.11487  (2022)

\bibitem{schuhmann2021laion}
Schuhmann, C., Vencu, R., Beaumont, R., Kaczmarczyk, R., Mullis, C., Katta, A.,
  Coombes, T., Jitsev, J., Komatsuzaki, A.: Laion-400m: Open dataset of
  clip-filtered 400 million image-text pairs. arXiv preprint arXiv:2111.02114
  (2021)

\bibitem{shen2017deep}
Shen, D., Wu, G., Suk, H.I.: Deep learning in medical image analysis. Annual
  review of biomedical engineering  \textbf{19}, ~221 (2017)

\bibitem{sohl2015deep}
Sohl-Dickstein, J., Weiss, E., Maheswaranathan, N., Ganguli, S.: Deep
  unsupervised learning using nonequilibrium thermodynamics. In: International
  Conference on Machine Learning. pp. 2256--2265. PMLR (2015)

\bibitem{song2020denoising}
Song, J., Meng, C., Ermon, S.: Denoising diffusion implicit models. arXiv
  preprint arXiv:2010.02502  (2020)

\bibitem{sudlow2015uk}
Sudlow, C., Gallacher, J., Allen, N., Beral, V., Burton, P., Danesh, J.,
  Downey, P., Elliott, P., Green, J., Landray, M., et~al.: Uk biobank: an open
  access resource for identifying the causes of a wide range of complex
  diseases of middle and old age. PLoS medicine  \textbf{12}(3),  e1001779
  (2015)

\bibitem{sun2022hierarchical}
Sun, L., Chen, J., Xu, Y., Gong, M., Yu, K., Batmanghelich, K.: Hierarchical
  amortized gan for 3d high resolution medical image synthesis. IEEE Journal of
  Biomedical and Health Informatics  (2022)

\bibitem{wang2020state}
Wang, L., Chen, W., Yang, W., Bi, F., Yu, F.R.: A state-of-the-art review on
  image synthesis with generative adversarial networks. IEEE Access
  \textbf{8},  63514--63537 (2020)

\bibitem{wang2021review}
Wang, T., Lei, Y., Fu, Y., Wynne, J.F., Curran, W.J., Liu, T., Yang, X.: A
  review on medical imaging synthesis using deep learning and its clinical
  applications. Journal of applied clinical medical physics  \textbf{22}(1),
  11--36 (2021)

\bibitem{zhang2018unreasonable}
Zhang, R., Isola, P., Efros, A.A., Shechtman, E., Wang, O.: The unreasonable
  effectiveness of deep features as a perceptual metric. In: Proceedings of the
  IEEE conference on computer vision and pattern recognition. pp. 586--595
  (2018)

\end{thebibliography}
\end{document}